\documentclass[twocolumn,showpacs,aps,pra]{revtex4-1}

\usepackage{amsmath,amsfonts,amssymb}
\usepackage{wrapfig}
\usepackage{graphicx}
\usepackage{bbm}
\usepackage{tabularx}
\usepackage{color}

\begin{document}

\title{Experimental verification of multidimensional quantum steering}\date{\today}

\author{Che-Ming Li$^{1}$}
\email{cmli@mail.ncku.edu.tw}
\author{Hsin-Pin Lo$^{2}$}
\email{hplo@nctu.edu.tw}
\author{Liang-Yu Chen$^{1}$}
\author{Atsushi Yabushita$^{2}$}
\email{yabushita@mail.nctu.edu.tw}

\affiliation{$^1$Department of Engineering Science, National Cheng Kung University, Tainan 701, Taiwan}
\affiliation{$^2$Department of Electrophysics, National Chiao-Tung University, Hsinchu City 300, Taiwan}

\begin{abstract}
Quantum steering enables one party to communicate with another remote party even if the sender is untrusted. Such characteristics of quantum systems not only provide direct applications to quantum information science, but are also conceptually important for distinguishing between quantum and classical resources. While concrete illustrations of steering have been shown in several experiments, quantum steering has not been certified for higher dimensional systems. Here, we introduce a simple method to experimentally certify two different kinds of quantum steering: Einstein-Podolsky-Rosen (EPR) steering and single-system (SS) steering (i.e., temporal steering), for dimensionality ($d$) up to $d=16$. The former reveals the steerability among bipartite systems, whereas the latter manifests itself in single quantum objects. We use multidimensional steering witnesses to verify EPR steering of polarization-entangled pairs and SS steering of single photons. The ratios between the measured witnesses and the maximum values achieved by classical mimicries are observed to increase with $d$ for both EPR and SS steering. The designed scenario offers a new method to study further the genuine multipartite steering of large dimensionality and potential uses in quantum information processing.
\end{abstract}

\maketitle

\section{Introduction}
Einstein-Podolsky-Rosen (EPR) steering, which was originally
introduced by Schr\"{o}dinger \cite{schrodinger1935discussion,schrodinger1936probability} in response to the EPR
paradox \cite{einstein1935can}, describes the ability of one party, Alice, to affect the state of another remote party, Bob, through her measurements on one of an entangled pair shared between them. Such effect recently is reformulated in terms of a information-theoretic task \cite{wiseman2007steering} showing that two parties
can share entanglement even if the measurement devices of Alice are
untrusted. This also shows a hierarchy between Bell non-locality, steering and entanglement. To rule out the classical mimicry of steering, several important methods are introduced to detect the steerability of bipartite quantum systems, for instance, the EPR steering inequalities \cite{cavalcanti2009experimental} and the steering measures \cite{skrzypczyk2014quantifying,piani2015necessary}. Combined with the tools to certify steering, EPR steering has stimulated application to quantum key distribution (QKD)
when one of the parties can not trust their measurement apparatus, i.e., one-sided device-independent QKD \cite{branciard2012one}.

There has been a range of investigations into potential extensions of EPR steering since the reformulation introduced by Wiseman, Jones and Doherty \cite{wiseman2007steering}. For example, it has been shown that EPR steering can occur in only one direction \cite{midgley2010asymmetric,olsen2013asymmetric,bowles2014one}, from Alice to Bob but not from Bob to Alice. Genuine multipartite steering \cite{he2013genuine} are introduced to generalize the original bipartite steering effect. In addition, a temporal analog of the steering inequality has been introduced \cite{chen2014temporal}. The concept of quantum steering for single quantum systems and its role in quantum information processing have been investigated further \cite{li2015certifying}. 

For practical tests of steering, experimental demonstrations of EPR steering have been presented in several quantum systems. These experiments successfully test steering among bipartite \cite{saunders2011experimental,smith2012conclusive,wittmann2012loophole,bennet2012arbitrarily,handchen2012observation,steinlechner2013strong,sun2014experimental} and multipartite \cite{armstrong2015multipartite,cavalcanti2015detection,li2015genuine,Deng17} systems. A detection of temporal steering also has been reported recently \cite{bartkiewicz2016temporal}. Inspired by all of these studies on steering between remote parties or temporal points, we go a step further and consider the following question: how does the quantum steering change with the dimension of considered systems? While the original concern of steering focus on the overall characteristic of a physical system that shows steerability, the system dimensionality indeed plays a role in manifesting properties that constitute a physical object. Here, we use the newly introduced quantum witnesses \cite{li2015certifying} to experimentally observe quantum steering. Both EPR steering and the single-system (SS) steering, i.e., temporal steering \cite{chen2014temporal,li2015certifying}, are considered in our experimental demonstrations. As will be shown presently, these steering effects vary with the system dimensionality and reveal stronger non-classical features as the dimensionality increases. The present study investigates further the utilities of the steering witnesses \cite{li2015certifying} and shows, to our knowledge, the first experimental demonstration revealing an increase of quantum steering with dimensions.

\begin{figure}[t]
\centering
\includegraphics[width=7cm]{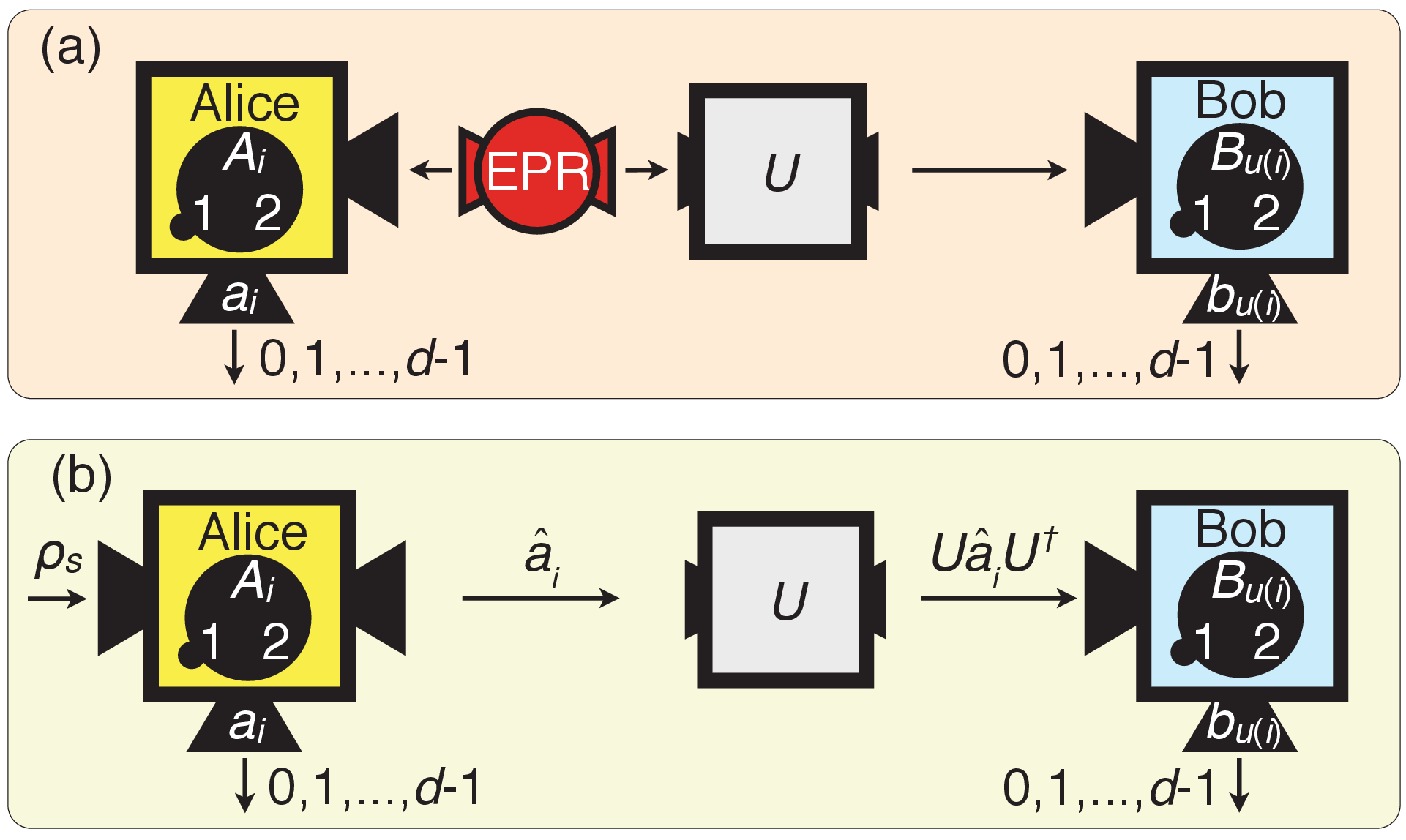}
\caption{Quantum steering. (a) EPR steering, and (b) sing-system steeering. In (a), for the ideal case, Alice can perform measurements on her qudit of an entangled pair generated from EPR source and implement the operation $U$ such that Bob's qudit state is in $U\hat{a}_{i}U^{\dag}$ where $\hat{a}_{i}\equiv \left|a_{i}\right\rangle_{ii\!}\left\langle a_{i}\right|$. See Eq.~(\ref{EPRu}) and its discussions. In (b), a qudit with the state $\hat{a}_{i}$ is sent from Alice to Bob. Here $\hat{a}_{i}$ is a post-measurement state of a initial qudit $\protect\rho _{S}$ under the measurement $A_{i}$ for $i=1,2$. Alice can steer the state of Bob's particle into other states by doing herself or asking him to perform the quantum operation $U$. While the resources utilized for quantum steering in (a) and (b) are different, the state of the particle finally held by Bob can be steered into a corresponding quantum state, $U\hat{a}_{i}U^{\dag }$, for both quantum steering scenarios.}\label{steering}
\end{figure}

In order to introduce our experimental scenario and the main results, we firstly present an unified way to review EPR steering and the SS steering. Alice's ability to affect the quantum state Bob has access to is based on (1) her ability to prepare a quantum source shared between her and Bob, and (2) her knowledge about the state Bob finally receives; see Fig.~\ref{steering}. Alice utilizes entangled pairs of quantum $d$-dimensional systems (qudits) as the quantum source when showing EPR steering [Fig.~\ref{steering}(a)], whereas she generates single-quantum systems with arbitrary states to Bob for the SS steering [Fig.~\ref{steering}(b)]. When Alice is certain that the entangled qudits eventually shared between them is as expected, she can prepare a target state for Bob by measuring her qudit of the entangled pair. Compared with such preparation of Bob's state in EPR steering, the target state in the scenario of SS steering is prepared by directly sending single systems with designed states from Alice to Bob. For both types of quantum steering, if Alice has full information about the quantum system Bob is holding, she is capable of steering the system into an arbitrary state.

\subsection{Einstein-Podolsky-Rosen steering}

Let us concretely show how Alice achieves EPR steering reviewed above. First, the entanglement source (or called EPR source) [Fig.~\ref{steering}(a)] creates $d$-dimensional entangled pairs of the form
\begin{equation}
\left|\Phi\right\rangle=\frac{1}{\sqrt{d}}\sum_{a_{1}=b_{1}=0}^{d-1}\left|a_{1}\right\rangle_{A1}\otimes\left|b_{1}\right\rangle_{B1}\label{Phi}
\end{equation}
where $\{\left|a_{1}(b_{1})\right\rangle_{A1(B1)}\equiv\left|a_{1}(b_{1})\right\rangle_{1} \mid a_{1}(b_{1})\in\textbf{v}=\{0,1,...,d-1\}\}$ is an orthonormal bases. $A_{1}$ and $B_{1}$ denote the measurement of Alice and Bob, respectively. Second, Alice keeps one particle of the entangled pair and sends the other qudit to Bob. A subsequent unitary operator $U$ is applied on the Bob's subsystem according to the instructions of Alice. This transformation can be done either by Bob after receiving the particle, or by Alice herself before the transmission of the particle. After such transformation, the state vector of the bipartite system becomes
\begin{eqnarray}
(I\otimes U)\left|\Phi\right\rangle=\frac{1}{\sqrt{d}}\sum_{a_{1}=b_{1}=0}^{d-1}\left|a_{1}\right\rangle_{A1}\otimes U\left|b_{1}\right\rangle_{B1}.\label{EPRu}
\end{eqnarray}
Then, depending on Alice's measurement result $a_{1}$, the state of the particle finally held by Bob can be steered into a corresponding quantum state, $U\hat{a}_{1}U^{\dag }$. When the state $\left|\Phi\right\rangle$ is represented in the bases $\{\left|a_{2}\right\rangle_{A2}\equiv\left|a_{2}\right\rangle_{2} \mid a_{2}\in\textbf{v}\}$ for the measurement $A_{2}$ and $\{\left|b_{2}\right\rangle_{B2}\equiv\left|b_{2}\right\rangle_{2} \mid b_{2}\in\textbf{v}\}$ for the measurement $B_{2}$, where $\left| k_{2}\right\rangle_{2} =1/\sqrt{d
}\sum_{k_{1}=0}^{d-1}\exp (i 2\pi k_{1}k_{2}/d)\left| k_{1}\right\rangle_{1}$ for $k=a,b$, we have
\begin{equation}
\left|\Phi\right\rangle=\frac{1}{\sqrt{d}}\sum_{a_{2}+b_{2}\doteq 0}\left|a_{2}\right\rangle_{A2}\otimes\left|b_{2}\right\rangle_{B2},\label{phi2}
\end{equation}
where $\doteq$ denotes equality modulo $d$. As Alice measures on her qudit with a result $a_{2}$, Bob's qudit is then steered into the quantum state, $U\hat{b}_{2}U^{\dag }$, where $a_{2}+b_{2}\doteq 0$. If Bob takes the complementary measurements on his particle $B_{u(1)}$ and $B_{u(2)}$ that are specified by the orthonormal bases $\{\left| b_{u(i)}\right\rangle_{u(i)} \equiv U\left|
b_{i}\right\rangle_{i} \mid b_{u(i)}=b_{i}\in \mathbf{v}\}$, he will know the results $\{b_{u(i)}\}$ designed by Alice with certainty.

We remark that, for an EPR source creating entangled states that are different from $\left|\Phi\right\rangle$, the transformation $U$ could be implemented in other ways. For example, when Alice and Bob share bipartite supersinglets \cite{cabello2002n}, which are expressed as $\left|\Psi\right\rangle=\frac{1}{\sqrt{d}}\sum_{a_{i}+b_{i}=d-1}(-1)^{a_{i}}\left|a_{i}\right\rangle_{Ai}\otimes\left|b_{i}\right\rangle_{Bi}$, for $i=1,2$, Alice can steer the state of Bob by directly measuring her qudit in a basis featured in $U$. Since supersinglets are rotationally invariant \cite{cabello2002n}, i.e., $(R\otimes R)\left|\Psi\right\rangle=\left|\Psi\right\rangle$, where $R$ is a rotation operator, Alice's measurement in the basis $\{R\left|a_{i}\right\rangle_{i}\}$ will steer the state of Bob's qudit into a corresponding state, $R\left|b_{i}\right\rangle_{i}$, for $a_{i}+b_{i}=d-1$. For $d=2$, supersinglets become unitary invarient and provide a resource for implementing any unitary transformations $U$ to Bob's qubit.

\subsection{Single-system steering}

Compared with EPR steering, Alice can realize the SS steering by following the single-system-analogue steps. As depicted in Fig.~\ref{steering}(b), first, Alice prepares a state $\hat{a}_{i}\equiv \left|a_{i}\right\rangle_{ii\!}\left\langle a_{i}\right|$ by performing complementary measurements $A_{1}$ or $A_{2}$ on an initial state, say $\rho _{S}$. Second, Alice sends the particle with the state $\hat{a}_{i}$ to Bob and steers the state $\hat{a}_{i}$ into other quantum state $U\hat{a}_{i}U^{\dag }$, by directly performing the unitary transformation $U$ before the particle transmission, or publicly, via a classical channel, ask Bob to apply $U$ on $\left|a_{i}\right\rangle_{i}$. Here the complementary measurements on Bob's particle $B_{u(1)}$ and $B_{u(2)}$ are specified by the orthonormal bases $\{\left| b_{u(i)}\right\rangle_{u(i)} \equiv U\left|
b_{i}\right\rangle_{i}\}$ with the results $\{b_{u(i)}\}$.

\section{Multidimensional steering witnesses}

The steering features of the entangled states $\left|\Phi\right\rangle$ and the states of single quantum systems $\hat{a}_{i}$ can be revealed by using steering witnesses. These tools considered in our experiments are of the from $\mathcal{W}>\alpha_{R}$ \cite{li2015certifying}, where $\mathcal{W}$ is the witness kernel and $\alpha_{R}$ is the maximum value of the kernel supported by classical mimicry. $\mathcal{W}$ are designed according to some target quantum sources, then such quantum witnesses can be experimentally implemented without invoking any quantum tomographic techniques. For ideal steering, $\mathcal{W}$ will be maximized. Since ruling out classical mimicry is equivalent to excluding unsteerable states, exceeding the bound $\alpha_{R}$ will deny processes (e.g., noisy channels) that make once steerable states unsteerable and thus assist in confirming genuine quantum steering.

The witness kernel of EPR steering used in our experimental verification is
\begin{equation}
\mathcal{W}_{dU,EPR}\equiv \sum_{a_{1}=0;b_{1}=a_{1}}^{d-1}P(a_{1},b_{1})+\sum_{a_{2}=0;a_{2}+b_{2}\doteq 0}^{d-1}P(a_{2},b_{2}),  \label{wepr}
\end{equation}
where $P(a_{i},b_{i})$ are the joint probabilities of getting $a_{i}$ by Alice and $b_{i}$ by Bob. For SS steering, the witness kernel reads
\begin{equation}
\mathcal{W}_{dU,	SS}\equiv \sum_{i=1}^{2}\sum_{a_{i}=0;b_{i}=a_{i}}^{d-1}P(a_{i},b_{i}).  \label{wss}
\end{equation}
The joint probability can be represented further by $P(a_{i},b_{i})=P(a_{i})P(b_{i}|a_{i})$, where $P(a_{i})$ and $P(b_{i}|a_{i})$ denote the marginal probability of measuring $a_{i}$ by Alice and the probability of obtaining $b_{i}$ by Bob conditioned on the Alice's result $a_{i}$, respectively. Here, without losing any generality, we have assumed that $U=I$ in the steering scenario (see Fig.~\ref{steering}). For ideal $d$-dimensional steering by using the source $\left|\Phi\right\rangle$ or $\hat{a}_{i}$, these kernels have the maximum value $\mathcal{W}_{dU}=2$ (note that we shall use $\mathcal{W}_{dU}$ to signify both $\mathcal{W}_{dU,EPR}$ and $\mathcal{W}_{dU,SS}$ hereafter). 
Whereas, for the
unsteerable states, the maximum value of $\mathcal{W}_{dU}$ is $\alpha _{R}=1+1/\sqrt{d}$ for both the types of steering witnesses. Thus, if experimental results show that
\begin{equation}
\mathcal{W}_{dU}>1+\frac{1}{\sqrt{d}},  \label{wdu}
\end{equation}
then the created states are steerable. For any unsteerable states the measured kernel will not certified by the witnesses \cite{li2015certifying}.

One of the important features of the steering witnesses considered in this paper is that $\mathcal{W}_{dU}<2$ for any $d'$-dimensional systems where $d'<d$. The reason is that the witness kernels $\mathcal{W}_{dU,EPR}$ (\ref{wepr}) and $\mathcal{W}_{dU,SS}$ (\ref{wss}) are composed of joint probabilities for outcomes observed under two complementary measurements such that the maximum of $\mathcal{W}_{dU}=2$ can not be satisfied by $d'$-dimensional systems. 

For the witness kernel $\mathcal{W}_{dU,EPR}$, it is clear that $\left|\Phi\right\rangle$ is the only state such that $\mathcal{W}_{dU,EPR}=2$. Hence, for any states with $d'<d$, the measured witness kernels are smaller than $2$. For the case of the SS steering, let us assume that Alice's apparatuses support desired measurements on $d$-dimensional systems, but, for state preparations, she has only the ability to create $d'$-dimensional states sent to Bob for $d'<d$. On the side of Bob, we assume that his measurement devices can realize state distinctions between $d$-dimensional states for $B_{1}$ and $B_{2}$ measurements. Suppose that $\rho_{S}(d')=\left|0\right\rangle_{11\!}\left\langle 0\right|$ of a $d'$-dimensional system is sent to Alice for measurements and that Bob's qudit state is the same as the qudit sent by Alice, we have the maximal sum of the joint probabilities under the measurement $(A_{1},B_{1})$, i.e., $\sum_{a_{1}=0;b_{1}=a_{1}}^{d-1}P(a_{1},b_{1})=1$. Since $(A_{2},B_{2})$ is complementary to $(A_{1},B_{1})$, the maximal sum Alice's $d'$-dimensional systems can give is only
\begin{eqnarray}
\sum_{a_{2}=0;b_{2}=a_{2}}^{d-1}P(a_{2},b_{2})&=&(\frac{1}{\sqrt{d}})^{2}(\frac{d'}{\sqrt{dd'}})^{2}d\nonumber\\
&=&\frac{d'}{d}.
\end{eqnarray}
The first term, $(1/\sqrt{d})^{2}$, is the probability $P(a_{2})$ and the last two are derived from the $d$ terms of the maximum of the overlap between the prepared state by Alice $\hat{a}_{2}(d')$ and the projected state $\hat{b}_{2}(d)$ of Bob's measurements. Thus for such $d'$-dimensional systems we have $\mathcal{W}_{dU,SS}=1+d'/d$.\\     


In oder to concretely see how steering effects vary with the system dimensionality, we consider the ratio between the measured witness and the maximum value achieved by classical mimicries of the form:
\begin{equation}
R_{d}:=\frac{\mathcal{W}_{dU}}{1+\frac{1}{\sqrt{d}}}.\label{Rd}
\end{equation}
For the cases where state preparations and measurements are perfect, i.e., we have the theoretical values $\mathcal{W}_{dU}=2$, it is clear to see that this ratio is monotonically increasing with $d$ and $R_{d_{1}}>R_{d_{2}}$ for any $d_{1}>d_{2}$, for example, 
\begin{equation}
R_{2}\simeq 1.1712,\
R_{4}\simeq 1.3333,\ 
R_{8}\simeq 1.4776,\ 
R_{16}\simeq 1.6000, \nonumber
\end{equation}
and for the case of large $d$, we have $R_{d}\simeq 2$. The steering properties certified by the witnesses~(\ref{wdu}) reveal that, compared with the usual quantum steering of qubits ($d=2$), multidimensional quantum systems ($d\geq 3$) provide stronger steering and manifest more distinct quantum violations as $d$ increases.

 It is interesting to compare this increasing trend with the results derived from Bell nonlocality of multidimensional systems. As shown by Mermin \cite{mermin1990extreme}, $N$ spin-1/2 particles in the Greenberger-Horne-Zeilinger (GHZ) state can possess correlation that violates a Bell inequality by an amount that increases exponentially with $N$. When $N$ is even, the corresponding increase by GHZ state is $R_{\text{M},even}=2^{(N-2)/2}$. If each half of the $N$ particles constitutes a qudit with $d=2^{N/2}$, then the factor can be rephrased as: $R_{\text{M},even}=d/2$, which is linearly increasing with the system dimension $d$. When using the Bell inequalities introduced by Collins, Gisin, Linden, Massar, and Popescu (CGLMP) \cite{collins2002bell} to consider the quantum-to-classical ratios, the state (\ref{Phi}) can show quantum violations with increasing ratios as well, for instance \cite{collins2002bell}: $R_{\text{CGLMP},2}\simeq 2.8729/2\simeq1.4365$, $R_{\text{CGLMP},3}\simeq 2.8962/2=1.4481$, and $R_{\text{CGLMP},d}\simeq 2.9696/2=1.4848$ for large $d$.

\begin{figure*}[t]
\centering
\includegraphics[width=12.2cm]{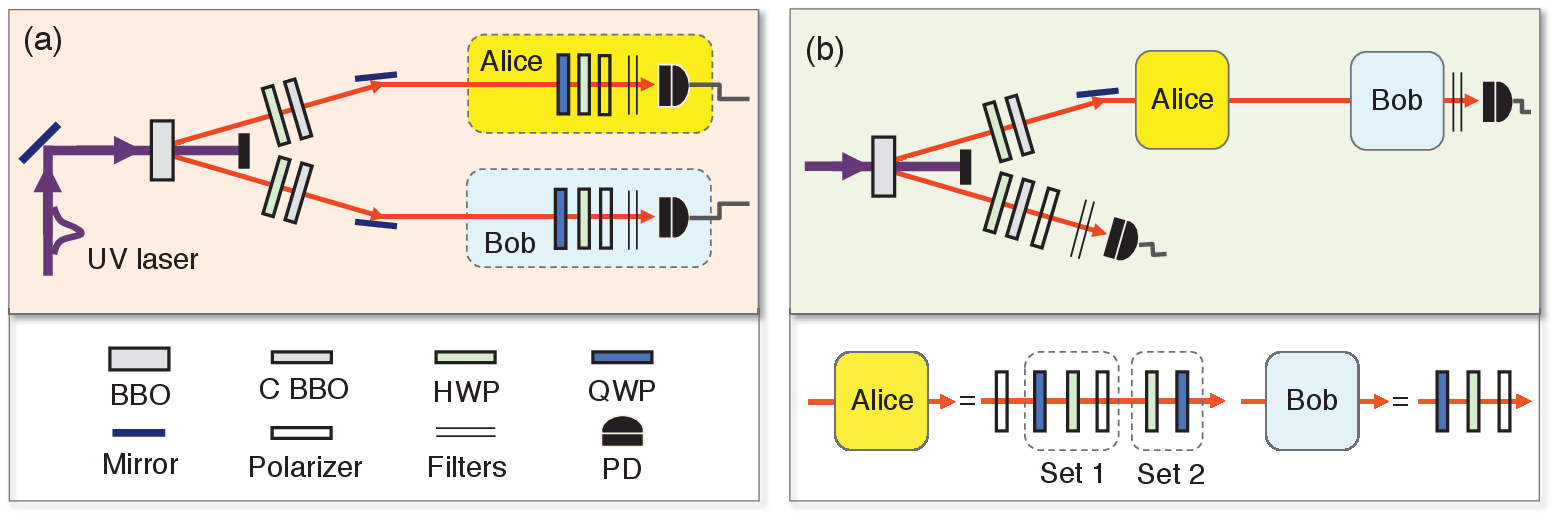}
\caption{Experimental set-up for testing multidimensional quantum steering. (a) Set-up for EPR steering. The ultraviolet (UV) pulsed laser (200 mW) is generated by second-harmonic generation with a Ti:Sapphire laser ($\lambda$=780 nm, pulse duration of 120 fs, and repetition rate of 76 MHz). The laser is used to pump the 2-mm-thick type-II $\beta$-barium borate (BBO) crystal to create polarization-entangled photon pairs $\rho_{\phi}$ by the spontaneous parametric down-conversion (SPDC) process. The 1-mm-thick C BBO and half-wave plate (HWP) are used to walk-off compensation. A HWP, a quarter-wave plate (QWP) and a polarizer on both sides of Alice and Bob are used to perform measurements on single-photon polarization states \cite{james2001measurement,lo2016experimental}. All photons are filtered by interference filters (Semrock: LL01-780-25, 3 nm) and are measured by a single-photon counting modules (Perkin-Elmer, SPCM-AQR-14). Coincidences are then recorded by a time-to-amplitude convert (ORTEC, model 567). (b) Set-up for the SS steering. One photon of an entangled pair is used to show the SS steering, and its initial state ($\rho_{s}=\left|0\right\rangle_{11}\left\langle 0\right|$ in Fig.~\ref{steering}) is prepared by conditionally projecting both photons of the entangled pair onto $\left|H\right\rangle$ by placing a polarizer in each photon path. The first set of wave plates (Set 1) is used to measure $P(a_{i})$ [see Eqs.~(\ref{wss}) and (\ref{pss})], whereas the second set (Set 2) prepares $\hat{a}_{i}$ sent to Bob. A wave-plate set which is conjugated to Set 2 on Bob side is used to measure $P(b_{i})$.}\label{setup}
\end{figure*}

\section{Experimental observation}

To investigate the characteristic of multidimensional steering predicted from the steering witnesses [Eqs. (\ref{wdu}) and (\ref{Rd})], we define that the qudit is composed of multiple particles. This method is also used in parallel to prepare multidimensional systems for testing Bell inequalities \cite{lo2016experimental}. Assuming that each particle is a two-state quantum object (qubit), the dimension of the Hilbert space of the ensemble consisting of $N$ qubits will be $d=2^{N}$. If we have $N$ entangled qubits, $\left|\phi\right\rangle_{m}=(\left|0\right\rangle_{A1,m}\otimes\left|0\right\rangle_{B1,m}+\left|1\right\rangle_{A1,m}\otimes\left|1\right\rangle_{B1,m})/\sqrt{2}$, for $m=1,2,...,N$, the total $N$-pair system is exactly a maximally entangled state of two qudit $\left|\Phi\right\rangle$ [Eq.~(\ref{Phi})] for EPR steering, i.e.,
\begin{eqnarray}
\left|\Phi\right\rangle&=&\bigotimes_{m=1}^{N}\frac{1}{\sqrt{2}}(\left|0\right\rangle_{A1,m}\otimes\left|0\right\rangle_{B1,m}+\left|1\right\rangle_{A1,m}\otimes\left|1\right\rangle_{B1,m})\nonumber\\
&=&\frac{1}{\sqrt{d}}\sum_{j=0}^{d-1}\left|j\right\rangle_{A1}\otimes\left|j\right\rangle_{B1},
\end{eqnarray}
where 
\begin{equation}
\left|j\right\rangle_{A1}=\bigotimes_{m=1}^{N}\left|j_{m}\right\rangle_{A1,m},\left|j\right\rangle_{B1}=\bigotimes_{m=1}^{N}\left|j_{m}\right\rangle_{B1,m}\label{s1}
\end{equation}
and $j=\sum_{m=1}^{N}j_{N-m+1}2^{m-1}$ for $j_{m}\in\{0,1\}$. Similarly, the state vectors $\left|k\right\rangle_{A2}$ and $\left|l\right\rangle_{B2}$ [Eq.~(\ref{phi2})] can be rephrased in terms of qubit states by
\begin{eqnarray}
\left|k\right\rangle_{A2}=\bigotimes_{m=1}^{N}\left|k_{m}\right\rangle_{A2,m},\left|l\right\rangle_{B2}=\bigotimes_{m=1}^{N}\left|l_{m}\right\rangle_{B2,m},\label{s2}
\end{eqnarray}
where
\begin{eqnarray}
\left|k_{m}\right\rangle_{A2,m}&=&\frac{1}{\sqrt{2}}(\left| 0\right\rangle_{A1,m} +e^{i \frac{2 \pi}{2^{m}} k} \left| 1\right\rangle_{A1,m}),\nonumber\\
\left|l_{m}\right\rangle_{B2,m}&=&\frac{1}{\sqrt{2}}(\left| 0\right\rangle_{B1,m} +e^{i \frac{2 \pi}{2^{m}}l} \left| 1\right\rangle_{B1,m}).\label{s2s}
\end{eqnarray}
With the state decompositions (\ref{s1}) and (\ref{s2}), the qudits with the states $\hat{a}_{i}$ used for the SS steering can be then prepared from qubits in the states $\left|j_{m}\right\rangle_{A1,m}$ and $\left|k_{m}\right\rangle_{A2,m}$ as well. Hence the measurements $(A_{1},B_{1})$ and $(A_{2},B_{2})$ required to experimentally determine the witness kernel $\mathcal{W}_{dU}$ can be performed on individual qubits. 

Suppose that we have an experimental state $\rho_{\Phi}(d)=\bigotimes_{m=1}^{N}\rho_{\phi m}$ where $\rho_{\phi m}$ is the state of the $m$th entangled pair, the joint probabilities in the kernel $\mathcal{W}_{dU,EPR}$ predicted by quantum mechanics can be represented in terms of the probabilities of individual pairs of entangled states by 
\begin{eqnarray}
P(a_{i}=k,b_{i}=l)&=&\prod_{m=1}^{N}\text{Tr}[\left|k_{m}\right\rangle_{Ai,mAi,m}\! \left\langle k_{m}\right|\nonumber\\
&&\ \ \ \ \ \ \otimes \left|l_{m}\right\rangle_{Bi,mBi,m}\! \left\langle l_{m}\right|\rho_{\phi m}].
\end{eqnarray}
The above form follows from the fact that the $d$-dimensional density matrix $\rho_{\Phi}(d)$ can be constructed by $N$-pair of qubits. Hence, the kernel of the multidimensional steering witness $\mathcal{W}_{dU,EPR}$ can be determined from the outcomes of measurements of individual entangled pairs. For the case of two ensembles of perfect entangled states $\left|\Phi\right\rangle$, $\mathcal{W}_{dU,EPR}$ is maximal, that is, $\mathcal{W}_{dU}=2$ for any $d$. Similarly, in the scenario of SS steering, assume that we generate a qudit with the state $\rho_{S}(d)=\bigotimes_{m=1}^{N}\rho_{s m}$, where $\rho_{s m}$ is the $m$th qubit which constitutes the qudit, then the kernel $\mathcal{W}_{dU,SS}$ can be measured by taking measurements on individual qubits by 
\begin{eqnarray}
&&P(a_{i}=k,b_{i}=l)\nonumber\\
&&=\prod_{m=1}^{N}\text{Tr}[\left|k_{m}\right\rangle_{Ai,mAi,m}\! \left\langle k_{m}\right|\rho_{s m}]\nonumber\\
&&\ \ \ \ \  \ \ \ \ \times \text{Tr}[ \left|l_{m}\right\rangle_{Bi,mBi,m}\! \left\langle l_{m}\right|\rho'_{s m|a_{1}=k}].\label{pss}
\end{eqnarray}
where $\rho'_{s m|a_{i}=k}$ is the state of qubit eventually held by Bob as Alice sends the qubit $\left|k_{m}\right\rangle_{Ai,m}$. 

\begin{figure}[t]
\centering
\includegraphics[width=8.5cm]{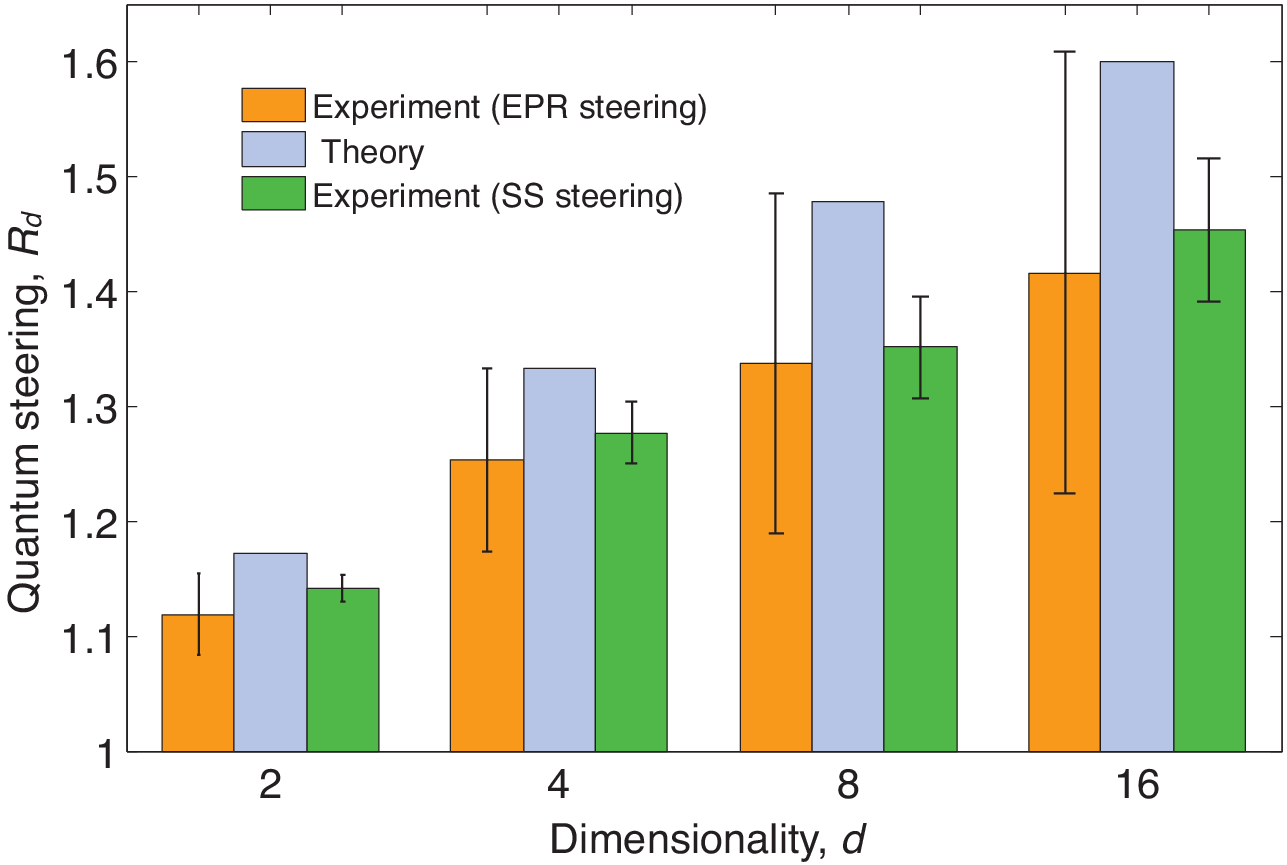}
\caption{Experimentally multidimensional quantum steering. Theoretical and experimental values of the ratio, $R_{d}$, between the witness kernel $\mathcal{W}_{dU}$ and the unsteerable bound $1+1/\sqrt{d}$ for $d=2,4,8,16$ are shown. Since all the experimental results indicate $R_{d}>1$, the prepared sources are identified as steerable. In particular, the experimental ratios increase with $d$ for both EPR and SS steering, which are consistent with the theoretical predictions of multidimensional quantum steering.}\label{results}
\end{figure}

In the experimental demonstrations of the above scheme, we use ensembles of photons to construct the qudits. For multidimensional EPR steering [Fig.~\ref{setup}(a)], the ingredient photon pairs are generated through the type-II SPDC (spontaneous parametric down-conversion) process and entangled at the degree of freedom of polarization in the form $\left|\phi\right\rangle_{m}=(\left|H\right\rangle_{A1,m}\otimes\left|H\right\rangle_{B1,m}+\left|V\right\rangle_{A1,m}\otimes\left|V\right\rangle_{B1,m})/\sqrt{2}$, where $\left|H(V)\right\rangle_{A1,m}$ and $\left|H(V)\right\rangle_{B1,m}$ represents the horizontal (vertical) polarization states of photons held by Alice and Bob, respectively. The methods of state preparation for certifying the SS steering are based on the same set-up for testing EPR steering [Fig.~\ref{setup}(b)]. To connect with the conceptual scheme, we make a correspondence of denotations by $\left|H\right\rangle\equiv \left|0\right\rangle$ and $\left|V\right\rangle\equiv\left|1\right\rangle$. In the experiment, the entangled pairs $\rho_{\phi m}$ consisting of $\rho_{\Phi}(d)$ and $\rho_{s m}$ consisting of $\rho_{S}(d)$ are created at different times. The stability of our laser and measurement system enables entangled pairs created at different times with a large time separation have a very close fidelity without additional system alinement, which makes the experimental states approximately identical at $\rho_{\phi m}\approx \rho_{\phi}$ for all pairs $m$ and the preparation of $\rho_{s m}$ more stable for steering tests. The entanglement source exhibits a high quality by the state fidelity $F_{\phi}=Tr[\rho_{\phi}\left|\phi\right\rangle \left\langle\phi\right|]\approx 0.982 \pm 0.006$, where $\left|\phi\right\rangle=(\left|H\right\rangle\otimes\left|H\right\rangle+\left|V\right\rangle\otimes\left|V\right\rangle)/\sqrt{2}$. It is worth noting that, while the system considered in our experiment is composed of subsystems created at different times, these subsystems eventually constitute a system with a $d$-dimensional state in the polarization degree of freedom. They can be locally measured to provide possible outcomes for determining the joint probabilities $P(a_{i},b_{i})$ and the kernels of $\mathcal{W}_{dU,EPR}$ and $\mathcal{W}_{dU,SS}$.

We use the same method as the approach presented in our work~\cite{lo2016experimental} on testing Bell inequalities to perform photon measurements for testing EPR steering. For SS steering, the qudits with the states $\hat{a}_{i}$ are described in the states $\left|j_{m}\right\rangle_{A1,m}$ and $\left|k_{m}\right\rangle_{A2,m}$ [Eqs.~(\ref{s1}) and (\ref{s2s})]. Their preparations and measurements then can be experimentally realized in the same way as shown above. See Figs.~\ref{steering}(b) and \ref{setup}(b). In our experiments, the initial qudit is prepared in the state $\protect\rho _{S}=\left|0\right\rangle_{11}\left\langle 0\right|$, i.e., the physical states of photons are created at $\left|0\right\rangle_{A1,m}=\left|H\right\rangle$ for all $m$'s. To realize this state preparation, we place a polarizer in each photon path to project both photons of the entangled pair onto state $\left|H\right\rangle$. The first wave-plate set (Set 1) is used to measure states of photons in the bases $\{\left|j_{m}\right\rangle_{A1,m}\}$ and $\{\left|k_{m}\right\rangle_{A2,m}\}$ for determining $P(a_{i})$ [see Eqs.~(\ref{wss}) and (\ref{pss})]. The subsequent wave-plate set (Set 2) is utilized to prepare specific polarization states ($\left|j_{m}\right\rangle_{A1,m}$ or $\left|k_{m}\right\rangle_{A2,m}$) which constitutes the state $\hat{a}_{i}$. Then, to measure $P(b_{i})$, we design a wave-plate set which is conjugated to Set 2 on the side of Bob.

Our experiment shows EPR steering and the SS steering for systems of up to $16$ dimensions. The kernel of multidimensional steering witness $\mathcal{W}_{dU}$ are calculated by measuring all of the probabilities $P(a_{i},b_{i})$. As seen in Fig.~\ref{results}, the experimental results are highly consistent with the theoretical predictions based on ideal entangled states $\left|\Phi\right\rangle$, state preparations $\hat{a}_{i}$ and perfect measurements. We clearly observe distinct quantum violations and their increases as the dimensionality $d$ raises. While the created states are close to the target state with high fidelity $F_{\phi}$, the witness kernel measured here are strictly dependent on the accurate settings of the wave plates. The total number of measurement settings of wave plates is $d-1$ for a given pair of operators $(A_{i}, B_{i})$. Then the required setting accuracy increases with $d$ proportionally. While arbitrary unitary transformations can be performed with high precision by sets of wave plates, such as the operations for single-photon polarization states (\ref{s2s}), imperfect angle settings can introduce errors that accumulate with increasing $d$. Therefore, compared to other noisy channels, such experimental imperfections become rather crucial in testing quantum steering of large dimensional systems.  See Ref.~\cite{rutkowski2017quantum} for detailed discussions about the issue of imprecise experimental verification and how to overcome the problem by including the tolerance for measurement-setting errors. It also can be analyzed in an manner similar to self-testing multipartite entangled states \cite{vsupic2017simple}. Although our demonstration shows cases up to $d=16$ only, the method can be straightforwardly extended to test multidimensional EPR steering and the SS steering for systems of larger $d$. 

\section{Conclusions}

In this paper, we have used multidimensional steering witnesses to experimentally certify EPR steering and the SS steering, for dimensionality up to $d=16$. In our experiment, polarization states of single photons and polarization-entangled photon pairs are utilized to serve as $d$-dimensional quantum sources in the scenarios of quantum steering. Such multidimensional systems show stronger steering effects than usual two-level objects. Compared with revealing this distinct feature by performing quantum information tasks, for example, by using two entangled pairs to teleport more than one-qubit information \cite{zhang2006experimental}, our experimental illustrations of steering give a new way to explicitly show the characteristics of multidimensional quantum sources. 

Our method to investigate multidimensional steering can be directly applied to genuinely multidimensional systems, for instance, the states of orbital angular momentum of photon pairs created through SPDC \cite{dada2011experimental}. Furthermore, it would be possible to study genuine multi-partite EPR steering \cite{he2013genuine} by extending the bipartite scenario presented in this work. A genuine eight-photon polarization-entangled state has recently been experimentally generated using the SPDC process \cite{yao2012observation}. With the recently introduced steering witnesses for genuine high-order EPR steering \cite{li2015genuine}, it holds high promise for observing EPR steering among the eight-partite high-dimensional systems. Since the witnesses for certifying genuine multi-partite EPR steering \cite{li2015genuine} has the same structure as the witness (\ref{wepr}) (see Eqs.~(1) and (6) in Ref.\cite{li2015genuine}), the increase of the classical-to-quantum ratio with dimension $d$ can still be seen, but the ratio is independent of the number of parties $N$. It is interesting to compare this case further with the quantum violations of Bell inequalities for genuine multi-partite Bell nonlocality \cite{brunner2014bell}. In addition to entangled photons, one can directly apply our idea to other quantum systems such as the multi-partite entangled ions in the GHZ state \cite{monz201114}.\\

\section*{Acknowledgements}
This work is supported by the Ministry of Science and Technology, Taiwan, under Grant No. MOST 104-2112-M-006-016-MY3 and 104-2112-M-009-001-MY2.


\end{document}